\documentclass[useAMS,usenatbib]{mn2e}
\usepackage{natbib}
\usepackage{graphicx}
\usepackage{subfigure}
\usepackage{url}

\title[Preparation of {\it Kepler} lightcurves for asteroseismic data analysis]{Preparation of {\it Kepler} lightcurves for asteroseismic analyses}
\author[]{R.A. Garc\'\i a,$^{1}$\thanks{E-mail: rgarcia@cea.fr(RAG); saskia@bison.ph.bham.ac.uk(SH)}  
S. Hekker,$^{2, 3}$ 
D.~Stello,$^4$
J. Guti\'errez-Soto,$^5$  
R.~Handberg,$^6$
D.~Huber,$^4$
\newauthor
C.~Karoff,$^6$
K.~Uytterhoeven,$^1$
T. Appourchaux,$^7$
W.~J.~Chaplin,$^3$
Y.~Elsworth,$^3$
\newauthor
S.~Mathur,$^{8}$
J.~Ballot,$^9$
J.~Christensen-Dalsgaard,$^6$
R.~L.~Gilliland,$^{10}$
G.~Houdek,$^{11}$
\newauthor
J.~M.~Jenkins,$^{12}$
H.~Kjeldsen,$^6$
S.~McCauliff,$^{13}$
T.~Metcalfe,$^{8}$
C.~K.~Middour,$^{13}$
\newauthor
J.~Molenda-Zakowicz,$^{14}$
M.~J.~P.~F.~G.~Monteiro, $^{15}$
J.~C.~Smith,$^{12}$
M.~J. Thompson,$^8$\\
$^{1}$Laboratoire AIM, CEA/DSM-CNRS-Universit\'e Paris Diderot; IRFU/SAp, Centre de Saclay, 91191 Gif-sur-Yvette Cedex, France\\
$^{2}$School of Physics and Astronomy, University of Birmingham, Edgbaston, Birmingham B15 2TT, United Kingdom\\
$^{3}$Astronomical Institute ``Anton Pannekoek'', University of Amsterdam, PO Box 94249, 1090 GE Amsterdam, The Netherlands\\
$^4$Sydney Institute for Astronomy (SIfA), School of Physics, University of Sydney, NSW 2006, Australia\\
$^{5}$Instituto de Astrof\'\i sica de Andaluc\'\i a (CSIC), Glorieta de la Astronom\'\i a s/n 18008, Granada, Spain\\
$^6$Department of Physics and Astronomy, Aarhus University, DK-8000 Aarhus C, Denmark\\
$^7$Institut d`Astrophysique Spatiale, Universit\'e Paris XI -- CNRS (UMR8617), Batiment 121, 91405 Orsay Cedex, France\\
$^8$High Altitude Observatory, National Center for Atmospheric Research, Boulder, Colorado 80307, USA\\
$^9$Laboratoire dÕAstrophysique de Toulouse-Tarbes, Universit\'e de Toulouse, CNRS, 31400 Toulouse, France\\
$^10$Space Telescope Science Institute, Baltimore, MD 21218, USA\\
$^{11}$Institute of Astronomy, University of Vienna, A-1180 Vienna, Austria\\
$^{12}$SETI Institute/NASA Ames Research Center, M/S 244-30, Moffett Field, CA 94035, USA\\
$^{13}$Orbital Sciences Corporation/NASA Ames Research Center, Moffett Field, CA 94035, USA\\
$^{14}$Astronomical Institute, University of Wroc{\l}aw, ul. Kopernika 11, 51-622 Wroclaw, Poland\\
$^{15}$Centro de Astrof\'\i sica, Universidade do Porto, Rua das Estrelas, 4150-762, Portugal
}

\begin{document}

\date{Accepted 2010 xxx xxx. Received 2010 October XX; in original form 2010 September 16}

\pagerange{\pageref{firstpage}--\pageref{lastpage}} \pubyear{2010}

\maketitle

\label{firstpage}

\begin{abstract}
The {\it Kepler mission} is providing photometric data of exquisite quality for the asteroseismic study of different classes of pulsating stars. These analyses place particular demands on the pre-processing of the data, over a range of timescales from minutes to months. Here, we describe processing procedures developed by the {\it Kepler} Asteroseismic Science Consortium (KASC) to prepare light curves that are optimized for the asteroseismic study of solar-like oscillating stars in which outliers, jumps and drifts are corrected.
\end{abstract}


\section{Introduction}

The primary scientific objective of the NASA {\it Kepler Mission} is to look for Earth-like planets in habitable zones of solar-like stars through the observation of photometric transits for at least 3.5 years  \citep{2010Sci...327..977B,2010ApJ...713L..79K}. Launched on March 7, 2009 (UTC), {\it Kepler} continuously monitors about 150,000 stars in a single field of view (FOV) of 115 $\rm deg^2$ located in the constellation of Cygnus that was selected to provide the optimal density of stars. 

{\it Kepler} is located in a 372.5-day, Earth-trailing, heliocentric orbit. This requires the satellite to perform 90$^{\circ}$ rolls about its axis every 93 days to keep the solar panels illuminated and the radiator, which cools the focal-plane arrays, pointed away from the Sun \citep{2010ApJ...713L.115H}. Data are consequently subdivided into quarters (denoted Qn or Qn.m, where n is the quarter number and m, the month), starting with the initial 10-day-long commissioning run (Q0), followed by a 34-day-long first quarter (Q1) and subsequent three-month-long quarters (Q2, Q3,...). 

\begin{figure*}
\includegraphics[trim =5mm 92mm 1mm 17.5mm, clip, height=0.195\textheight, width = 0.90\textwidth]{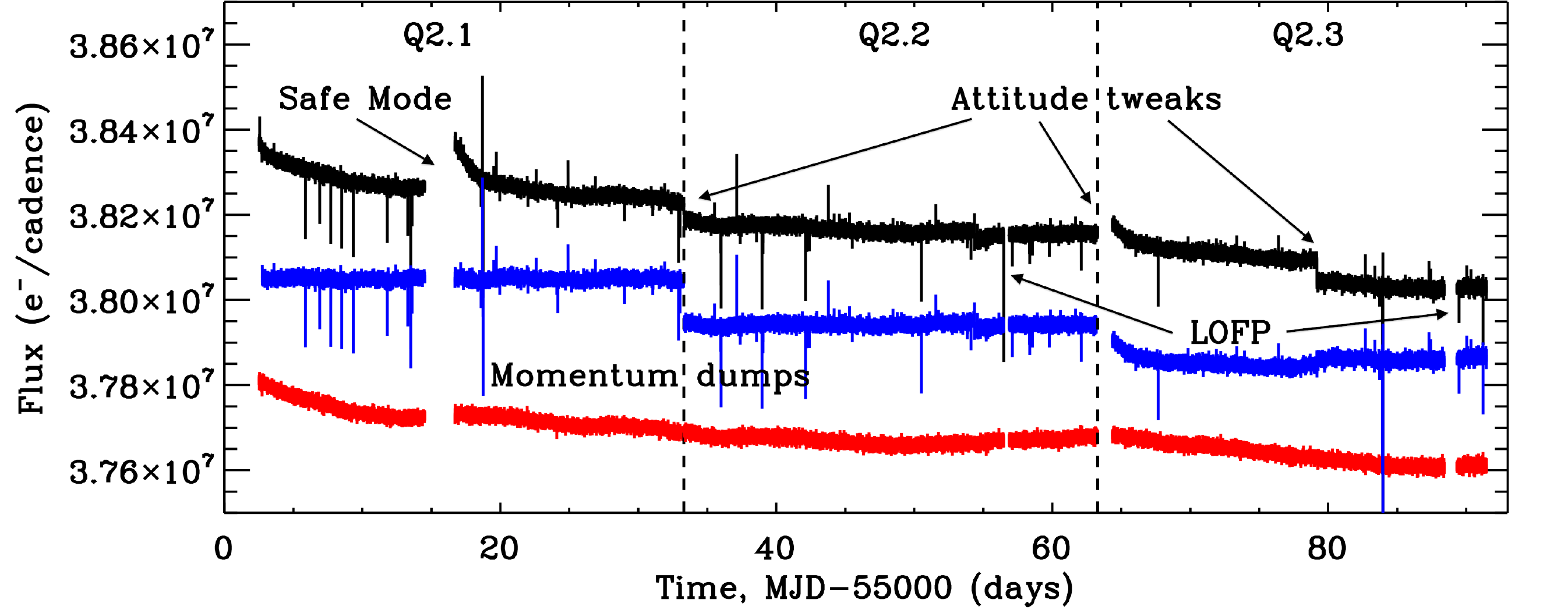}
\caption{\label{Q2} {Raw (black), PDC corrected (blue) and corrected --using the procedure described in this paper (red)-- light curves of the solar-like target: KIC~11395018 \citep{Mat11}. The corrected light curve has been shifted down, by $ 4\; 10^5 \;\rm{e^-/cadence}$, for the clarity of the comparison. The origin of the time axis is in Modified Julian dates (MJD) - 55000.  The points in which the fluxes fall abruptly are mostly due to} momentum-dump operations. LOFP stands for: Loss Of Fine Pointing.}
\end{figure*}

 The high precision of the differential photometry carried out by {\it Kepler} makes it an ideal instrument to perform asteroseismic studies --in which long and continuous observations are needed-- as part of the {\it Kepler} Asteroseismic Investigation \citep[KAI,][]{2010PASP..122..131G,KjeldsenJCD10}. Thus, the study of the resonant modes propagating inside a star complements the main scientific objectives of the {\it Kepler Mission} by characterizing the stars which potentially host planets \citep[e.g.][]{2010MNRAS.406..566M,2010ApJ...713L.164C,Gaulme2010} and their influence on the habitable zones, for example, due to their magnetic activity \citep[e.g.][]{2005Mosser,2009Mosser,2009MNRAS.399..914K,2010A&A...518A..53M,2010Sci...329.1032G}. Indeed, the properties of the eigenmodes depend on the internal structure and dynamics of the star in such a way that the fundamental stellar properties, such as their masses, radii, and ages,  can be inferred directly to levels that would be difficult to obtain by other more classical methods \citep[e.g.][Creevey et al. in preparation]{2009ApJ...700.1589S,2010ApJ...723.1583M,2010A&A...509A..77K}. Moreover, the unprecedented number of stars showing oscillations (covering most of the HR diagram), that are being observed by {\it Kepler} \citep[e.g.][]{2010ApJ...713L.176B,2010ApJ...713L.169C,2010ApJ...713L.192G,2010ApJ...713L.182S} as well as the French-led Convection Rotation and planetary Transits (CoRoT) satellite \citep[e.g.][]{2009A&A...506..465H,2009A&A...506...41G,2010A&A...515A..87D}, will soon modify our view of stellar evolution through the new constraints that we will be able to impose on the physical processes occurring in their interiors  \citep[e.g.][]{2009A&A...503L..21M,2010Miglio,Chap11}.
 
For each star, two types of light curves are available to the {\it Kepler} Asteroseismic Science Consortium (KASC) for asteroseismic investigations through the {\it Kepler} Asteroseismic Science Operations Center (KASOC) database (\url{http://kasoc.phys.au.dk/}):
on one hand, raw time series suffering from some instrumental perturbations; on the other hand, corrected light curves in which housekeeping data have been used to minimize those instrumental perturbations.  These second data sets are produced during the Pre-search Data Conditioning (PDC), enabling the search for exoplanet transits  \citep[][]{2010ApJ...713L..87J}. 
While these PDC datasets are in a constant evolution and new and more refined procedures are established, we found that, in some cases, part of the low-frequency stellar signal (such as the one produced by long-lived starspots \citep{Croll06}, short-lived starspots \citep{Mosser09} or low frequency modes) could be modified. Therefore, for solar-like oscillating stars as well as some classical pulsators ($\delta$-Scuti and $\Gamma$-Doradus stars), we decided to take the raw datasets and develop our own methods to correct for these perturbations.


\section{{\it Kepler} observations and stellar light curves}

{\it Kepler} observations are made in two different operating modes. Long cadence (LC) targets are sampled every 29.4244 minutes (Nyquist frequency of 283.45~$\mu$Hz) including all targets for exoplanets research for which signatures of photometric transits are sought. For the brightest stars (down to {\it Kepler} magnitude, $Kp \approx12$), short cadence (SC) observations can be obtained with a faster sampling rate of 58.84876~s (Nyquist frequency of $\sim$ 8.5~mHz), allowing for more precise transit timing. However, due to telemetry limitations, this running mode is only available for 512 targets. In both cases, the integration time is set to 6.02~s with a readout time of 0.52~s. The time is stamped in a way that the mid-time of each cadence is known with an accuracy of $\pm$ 0.050~s \citep[][]{2010ApJ...713L.160G}. Verification of this intended accuracy of the timing has not yet been done.

The light curves of the 150,000 stars are quasi-continuously recorded and stored on board the spacecraft. However, this data acquisition suffers from episodic data breaks due to operational procedures. Apart from the aforementioned spacecraft rolling, data acquisition is also interrupted once each month to download the stored science and engineering information. Indeed, the spacecraft has to be reoriented to point its high-gain antenna towards the Earth stopping the scientific data collection. Finally, every three days, one or more reaction wheels approach their maximum operating angular velocity. In order to ÒdesaturateÓ the wheels, the spacecraft fires its thrusters losing its attitude precise pointing for a few minutes. Once this operation is finished, {\it Kepler} returns to its normal fine-pointing mode but several targets could suffer from degraded pointing performance during these events. In general, about one data point in LC and several data points in SC are affected during each desaturation. Another problem related to the momentum management is when one of the reaction wheels crosses zero angular velocity. When this happens, the affected wheel rumbles and degrades the pointing on timescales of a few minutes. The primary consequence is an increased noise level in the SC centroid time series, with a resulting increase in noise in the pixel and flux time series.
 \citep[a more detailed information can be found in][]{2010ApJ...713L.115H}. Figure~\ref{Q2} shows a typical example of the Q2 raw light curve of a solar-like star (black), the PDC-corrected (blue) and the one we have corrected (red), in which all these interruptions can be seen.

The nominal timeline of {\it Kepler} science data collection has also been interrupted a few times due to unexpected events such as attitude tweaks, pointing drifts, periods of spacecraft jitter excess, loss of fine pointing (LOFP) events, etc. The most important interruptions - as judged by their lengths - are the so-called safe-mode events. This special mode tends to protect the spacecraft after an unanticipated response to a bad sequence of commands  or due to a problem in the on-board electronics after being hit by cosmic radiation. As an example, the longest safe-mode event occurred during the fourth roll (Q4) and lasted almost four days. During the second roll, a safe-mode event induced a two-day interruption (see Fig.~\ref{Q2}). The data collected after resumption of science data collection show a trend, which is strongly correlated with the warm-up of the local detector electronics. Depending on the CCD in which a star is observed, this effect manifests itself as a rising or decreasing trend in the light curve lasting for a few days (see Fig~\ref{Q2}).

\section{Correcting the instrumental perturbations}
To correct the raw light curves we follow a phenomenological approach in which we correct three types of effects: outliers, jumps, and drifts (no matter their physical origin). By doing so, we try to preserve, as much as possible, the low-frequency signal present in the data keeping in mind that some thermal or other long-term instrumental effects could still remain in the light curves. However, in the case of KASC Working Group 2 (``Stars in clusters''), we have compared all the stars corrected with this pipeline with the standard procedure \citep[PDC light curves,][]{2010ApJ...713L..87J}, adopting the corrections that performed best for every single target. Finally, we do not take into account the fraction of light thought to come from nearby stars. Therefore, in some cases, the amplitudes will be diluted.

\subsection{Outliers}
We have considered as outliers in the datasets the individual measurements showing a point-to-point deviation in the two-point difference function of the light curve greater than $3\sigma$ for the SC and $5\sigma$ for the LC, where $\sigma$ is defined as the standard deviation of the two-point difference function. Most of the points affected by this clipping are those {observed} during momentum desaturation maneuvers as well as during periods when the reaction wheels cross zero angular velocity. This correction also removes points affected by the {\it Argabrightening} effect, named after its discover by V. Argabright \citep{VanCleve_2009}. These points have an amplitude of many standard deviations above the average and its origin is not completely understood yet. It seems that they might be due to small dust particles from {\it Kepler} achieving escape velocity after micrometeorite hits and reflecting sunlight into the barrel of the telescope as they drift across the FOV \citep[further explanations can be found in][]{2010ApJ...713L..87J}. The outlier correction removes about one percent of the data points. The deleted points are written in the data file as ``-Inf'' (see also appendix A). In Fig.~\ref{Q2} we can see how all the outlier points seen in the raw light curve (black) were removed in the corrected one (red curve).

\subsection{Jumps}
Jumps are defined as sudden changes in the mean value of the light curve due to, for example, attitude tweaks or sudden drops in pixel sensitivity. The light curves have been checked at every cadence for these sudden changes by comparing the mean flux of one-day-long segments. When the difference between two adjacent segments is larger than a certain threshold, an additive correction is applied (multiplicative for the red-giants working group), i.e., adding or subtracting the difference in the average levels of the light curve segments with respect to the first part of the light curve. The threshold has been defined as five times the difference of the mean flux values of adjacent segments. Note that we always check for jumps at known times of satellite attitude changes as noted by \cite{VanCleve_2009}. 

The definition of the one-day segments in the jump correction only allows corrections in the light curve from the second day of measurements till the penultimate day of the run. There are no automatic algorithms in place to detect and correct for jumps in the first and last day of the quarter. These parts of the time series are inspected by eye and manually corrected if necessary.

An example of these corrections is shown in Fig.~\ref{Q2} in which 3 attitude tweaks produced a discontinuity in the   light curve, two of them (the first and third) were flagged as jumps and corrected. The second attitude tweak was considered as a drift (see the next subsection).
\begin{figure*}
\includegraphics[trim =5mm 92.5mm 1mm 16.3mm, clip, height=0.20\textheight, width = 0.90\textwidth]{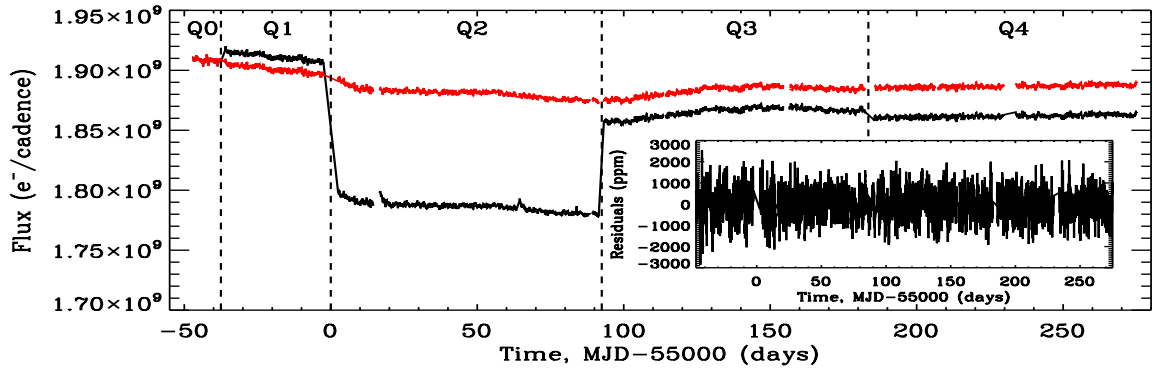}
\caption{\label{RG} {Raw (black) and corrected (red) light curves of the red giant target: KIC~1161618. The origin of the time axis is in Modified Julian dates (MJD) - 55000. The inset plot shows the residual light curve after applying the 10 days filter in the post-processing phase. The vertical dashed lines separate the different quarters.}}
\end{figure*}
\subsection{Drifts}

Drifts are small low-frequency perturbations, which are in general due to temperature changes (e.g., after a safe-mode event) and lasting for a few days. This correction is based on the software developed to correct the high-voltage perturbations in the GOLF/SoHO instrument \citep{GarSTC2005}. It consists of fitting a 2nd or 3rd order polynomial function to the region where a thermal drift has been observed after comparing several light curves of the same roll. Then, the fitted polynomial is subtracted in the affected portion of the light curve and we add another polynomial function (1st or 2nd order) ---used as a reference--- fitted to a local non-perturbed region of the time series which includes data before and after the discontinuity.  If the correction has to be applied on one border of the time series, only one side of the light curve is used to compute the reference function. After the correction, there is a manual validation of the result.


In Fig.~\ref{Q2} we can see this type of correction applied to the first few days of the quarter (Q2.1), during the safe mode event that also occurred in Q2.1, as well as during the second attitude tweak in which only the second segment (the beginning of Q2.3 data) has been modified.

\section{Merging light curves observed in more than one quarter}

Once the corrections are applied to the raw data, a single combined light curve has been constructed for each star observed for a period longer than a month. The mean flux levels of a star can differ considerably from quarter to quarter, because after rolling the satellite, stars are observed using another CCD module with different characteristics (aperture, crowding metrics, etc). Moreover, sometimes changes in the average flux inside a quarter can also happen. In the cases reported up to now, this problem is due to a bad definition of the optimal aperture for bright targets that saturate three or more pixels \citep[{\it Kepler} magnitude $Kp \le 11$,][]{VanCleve_2009}. With the progress of the mission most of the stars showing these aperture problems have been identified and their apertures improved, thereby minimizing this effect.

To implement the corrections at quarter boundaries, the first quarter for which data have been obtained for a given star is used as a reference. Depending on the processing of SC or LC time series, a different approach has been followed.

In the case of the SC data of KASC Working Group 1 (``Solar-like oscillating stars''), only a few stars have been observed longer than a month because it has been decided to perform a survey during the first year of {\it Kepler} scientific operations. During this survey, stars would be observed for a month. Thus, only six solar-like stars showing a p-mode hump were observed since the beginning of the mission \citep[four of them are deeply analyzed in][]{Camp11,Mat11}. To merge these data sets, we computed the mean value of segments of the light curve (each only one day long) at the start and end of each quarter, and we corrected for the difference using Q0 data as the reference. The resultant light curve showed a smoothed junction between all the quarters in the six stars. 

For LC data the correction procedure is slightly different. As we have processed more than a thousand stars in  KASC Working Group 8 (``Red giants'') and several hundreds (630 up to Q4) in Working Groups 4 and 10 (``delta-Scuti'' and ``gamma Doradus stars'', respectively), we took into account that, in some cases, there was a slope at the quarter edges due to temperature gradients. Therefore, we computed the mean flux values of the last two days of a quarter and the first two days of the next one as well as first order polynomial fits through the same segments of the light curve. Then, we added (or subtracted) the mean difference in flux to the second light curve based on either the mean values or the polynomials. Finally, for both solutions the light curves were merged and a polynomial was fitted through the four day segment of the merged light curve spanning the last two days of the first quarter and the first two days of the second one. The solution with the lowest $\chi^2$ was used for the final merged light curve. An example is shown in Fig.~\ref{RG} for the red-giant target KIC~1161618. While some long periods are still present in the light curve, the discontinuities at the quarter edges disappears.

The final merged light curve is cut back into the individual quarters and saved as an extension of the original files available at KASOC (see Appendix~A).

\section{Post-processing of remaining low-frequency signatures in the light curves}

Although we made an effort to correct for most of the instrumental effects and to retain the long-term (rotation/granulation signal) and short-term (oscillations) features in the light curves, we suspect that some instrumental effects remain in the corrected merged light curves (see also Appendix~B). Therefore, in some cases an additional filter should be applied to take into account these instrumental effects. The details of this filtering can differ for different types of stars and different scientific aims. As an example, in the case of red giants, we chose to perform an additional triangular smoothing with a filter 10 days wide (equivalent to 2 passes of rectangular smoothing) on
 the light curves. To avoid the influence of gaps on this smoothing, we apply it to an interpolated light curve into a regular grid of points and then we recompute the filtered signal into the {\it Kepler} timing. This smoothing removes the signals with timescales longer than 10 days, such as trends due to CCD degradation (see the inset plot in Fig.~\ref{RG}). This means that we can investigate the granulation, i.e., the signature of large convection cells present in the turbulent outer atmosphere of low-mass main-sequence stars and red giants, up to a timescale of 10 days (Mathur et al. in preparation, Mosser et al. in preparation). After removing the granulation signature, oscillations can be investigated. These have timescales of minutes for main-sequence stars up to hours or even a few days in red giants.

\section{Known artifacts in the PSD}

The power spectrum of the {\it Kepler} data suffers from some perturbations at given frequencies (mostly on SC data). The first spurious peaks to report are the harmonics of the 1/LC frequency at multiples of 566.4 $\mu$Hz that appears in the SC spectrum. The effect is bigger (more harmonics) for fainter stars, and it seems to be produced by the built-in electronics \citep{2010ApJ...713L.160G}.
A second group of peaks appears at constant frequencies at  very high frequency in SC spectra around 7024, 7444, 7865, and 8286 $\mu$Hz, which are separated by 421 $\mu$Hz (40 minutes). Finally, another two appears at 5017 and 5584 $\mu$Hz. All of them remain of unknown origin.

In the data obtained up to the second quarter, there are some peak-structures in the range 80 to 95 $\mu$Hz that are similar to some asteroseismic signatures due to the non-sinusoidal nature of the perturbation \citep[e.g.][]{2010arXiv1008.2959H}. These peaks are related to the variation of the reaction wheel housing temperature. They have been eliminated by reducing the corresponding temperature controller deadband \citep{VanCleve_2009}. There is also another peak-structure around 200 to 400 $\mu$Hz with associated artifacts between $\sim$500 to 530 $\mu$Hz that shift in frequency with time, for a yet unknown reason (Antochi, private communication).

Finally, spurious peaks have been found at 3 days, related to the momentum management cycle and its associated temperatures, and around 4500 $\mu$Hz, occurring mostly for stars with a moderate activity signal.

\section{Conclusions}
In this work we have described the corrections applied to the {\it Kepler} asteroseismic raw light curves to minimize some known instrumental effects and to produce the working group corrected files (labeled ``\_wg\#'' in the KASOC database, where ``$\#$'' denotes for the working group number). Three main phenomenological effects have been treated: outliers, jumps and drifts. In the case of time series longer than a month, a correction has been applied to smooth the discontinuities at the edges of the quarters. Finally, we have explained the structure of the data files and the known spurious frequencies that appear in the power spectrum of the asteroseismic targets.

\appendix
\section{File structure of  working group corrected data}
The corrected light curves are saved in an ASCII file with the same name as the original one but adding ``$\_$wg\#''', and uploaded into the KASOC database (a complete description can be found in Handberg \& Kjeldsen, KASC User Requirements Specifications: Working Group Corrected Data).

The structure of the file is the same as the original one but adding two extra columns containing the working group (wg) corrected flux and the wg corrected errors. For the moment, this last column is just a copy of the raw flux error. The points flagged as outliers are written as ``-Inf''.
 
The new file header contains some new information. A) An extra line is added below the first one to identify the file as being processed by the working group. This second line is of the form: ``\# Working Group N Corrected data by $<$Name$>$'' where N is the number of the working group and $<$Name$>$ is the name of the person who processed the new data file. B) The version number is changed adding a ``.YY'' where YY denotes the version number of the correction software (currently version 1). We also add the creation date of the file in parenthesis after a blank space. Thus, the line looks like: ``\# Version:    2.1 (1 Oct 2010)''. C) In the line describing the columns of the file we add: ``
WG\# Corrected Flux, WG \# Corrected error''. 

\section*{Acknowledgments}
Funding for this Discovery mission is provided by NASA's Science Mission. The authors wish to thank the entire {\it Kepler} team, all funding councils and agencies that have supported the activities of KASC, and the International Space Science Institute (ISSI). RAG wants to thank the support of the French PNPS program and SH the support from the UK Science and Technology Facilities Council (STFC) and the Netherlands Organization for Scientific Research (NWO).

\bsp

\label{lastpage}

\end{document}